\documentstyle[prl,aps,epsf,psfig,twocolumn,amstex,cite]{revtex}
\begin{document}
\def\lasco { La$_{2-x}$Sr$_x$CuO$_4$ }
\title{Jahn-Teller Impurity States in LaSrCuO: XAFS Evidence and
Implications for
High T$_c$ Superconductivity}
\author {Victor Polinger, Daniel Haskel and Edward A. Stern}
\address {Physics Department, Box 351560, University of
  Washington, Seattle  Washington 98195, USA}

\date{\today}
\maketitle

\begin{abstract}

Polarized XAFS measurements on powder of \lasco with aligned $c$-axes
find that apical oxygens neighboring only the Sr dopants have a
double-site distribution. This result requires that the doped holes
reside in impurity states peaked on the CuO$_6$ octahedron neighboring
the dopant (denoted as Sr-octahedrons). A model of the double site is
presented of two co-existing spin differentiated (singlet and triplet)
Jahn-Teller (JT) distortions of the Sr-octahedrons induced by an
extrinsic doped hole pairing with the intrinsic hole.  It is
speculated that Bose-condensation of the singlet pairs, bound by
$\gtrsim 0.1$ eV, produces superconductivity.
  
\end{abstract}


\narrowtext

Pure La$_2$CuO$_4$ is an insulating, strongly correlated electron
system with one intrinsic hole per molecular unit.  Doping with Sr at
the La site causes an insulator-to-metal transition (IMT) and a high
T$_c$ superconductor, \lasco.  There still remains controversy about
the mechanism of pairing that leads to superconductivity. We discuss
here implications of the recent XAFS experimental results that there
is a double-site distribution of the apical oxygens neighboring only
Sr dopant atoms\cite{Haskel-rapid}. Using compelling physical
arguments we arrive at a model which can explain the double-site
distribution and is consistent with other experimental and theoretical
published results that directly relate to the local atomic and carrier
properties of \lasco.  Possible implications for the mechanism of high
Tc superconductivity in this material are discussed.
    
Diffraction measurements\cite{Radaelli} have given detailed
information on the long range averaged periodic structure of \lasco.
However, since the superconducting coherence length of high T$_c$
materials is of the order of 10$\rm\, \AA$, the pertinent structure to
be used to understand superconducting properties is the local one over
this scale.  Whenever disorder is present in a crystalline material,
as occurs in \lasco because of the random Sr doping, the local and
average structure differ. XAFS\cite{stern} is the premier method to
determine the local structure up to 4-5$\rm\,
\AA$. Whereas diffraction techniques average over the La/Sr sites,
XAFS allows determining the structure separately about the La and Sr
atoms and, thus, can determine any perturbation introduced by the Sr
atoms.
    
The average structure of \lasco consists of single CuO$_2$ planes
separated by two La/Sr-O planes. The Cu atoms have four nearest
neighbor oxygens in the plane and two oxygens, called apicals, above
and below the plane, the six oxygens forming an elongated
octahedron. The apical oxygens, residing in the La/Sr-O planes, serve
an important function by forming the bond that allows transfer of the
doped holes to the CuO$_2$ planes. It is believed that the
mobile doped holes reside in the CuO$_2$ planes, produce metallic
conduction and are the carriers that are involved in the pairing
mechanism.
    
XAFS measurements were made on samples from the same batch used in the
neutron-diffraction study\cite{Radaelli} to ensure that any
differences between local and average structures are not due to
material differences. Quite complete information on the local
structures about both La and Sr atoms was obtained from our
data\cite{Haskel-rapid,Haskel-prl}, and new measurements at the Sr
$K$-edge extend the $x$-range far beyond the $x=0.075,0.10$ values
previously reported\cite{Haskel-rapid}. We concentrate here on the
results found for the apical oxygens as this information is most
pertinent for understanding the character of the carriers in this
material.
    
The apical oxygens about the La atoms are found to have a similar
distribution as in the pure material, namely, a single site located at
a distance of 2.35(1)$\rm\, \AA$ from the La atom (2.42(1)$\rm\, \AA$
from the almost collinear Cu atom). We denote such CuO$_6$ octahedrons
with only La atoms bonded to its two apical oxygens as
La-octahedrons. The apical oxygens about the Sr impurity atoms,
however, have a double-site distribution peaked at distances of
2.55(2)$\rm\, \AA$ and 2.25(3)$\rm\, \AA$ from Sr (2.22(2)$\rm\, \AA$
and 2.52(3)$\rm\,
\AA$ from the Cu atom, respectively), with the sum of occupation of the two
sites being one oxygen\cite{Haskel-rapid}. No change in both the
occupation and disorder of each site is found up to room temperature
(RT). However, the occupation of the sites varies with $x$, as shown
in Figure~1, with the long Sr-O distance becoming more occupied at the
expense of the short distance. No measurable variation of the Sr-O
apical distances occurs with $x$.

The significance of this result is its evidence of a strong
lattice-hole interaction only around the Sr atoms, indicating that
dopant-induced (extrinsic) holes are peaked only in the vicinity of the
CuO$_6$ octahedrons that are bonded to a Sr atom (the Sr-octahedrons).
The present result is different from the splitting in the apical
oxygen site reported in other investigations of high T$_c$
superconductors\cite{Mustre}, where the splitting occurs periodically
throughout the solid.  In our case, the double site occurs {\em only}
near the dopant Sr atoms.

\begin{figure}
\epsfxsize=2.5truein \epsffile{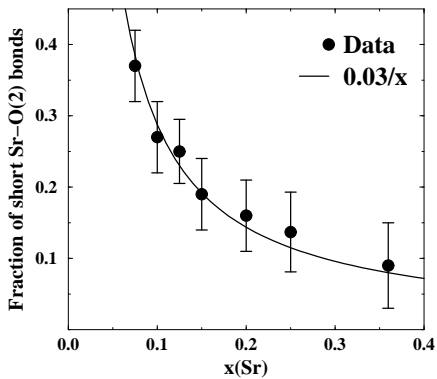} \vspace{0.1truein}
\caption{Fractional occupation of the short Sr-O(2)
distance as a function of Sr concentration $x$ obtained from Sr
$K$-edge XAFS measurements.}
\end{figure}

The Sr-octahedrons have two holes, one intrinsic and one extrinsic,
while the La-octahedrons have only an intrinsic hole.  Thus, the
extrinsic holes produce an impurity wave function distribution. From the
experimental result that doped
\lasco remains insulating up to a critical value of $x_c\sim 0.07$,
where it becomes metallic and superconducting, it follows that in the
dilute limit the extrinsic holes are immobilized about their
Sr-octahedrons till the impurity band broadening is sufficient to
overcome the localization energy. Thus, overlap is small and a
tight-binding approximation can describe the impurity states of the
extrinsic holes residing at Sr-octahedrons.
    
The energy states of CuO$_6$ octahedra can be used to gain insight
into the impurity states that occur in the solid.  We start with
one-electron energy levels, then add multi-electron (lattice-electron
and exchange interactions). The La-octahedrons, also in the undoped
material, have the electron energy levels shown in Fig.~2(a).  The
intrinsic hole is in the highest energy orbital with $x^2 - y^2$
symmetry while the lower lying $3z^2 - r^2$ orbital ($z^2$) is fully
occupied.  Even though these states have the symmetry of Cu {\it
d}-orbitals, they reside not only on the Cu atoms but on the
octahedron with a majority of the hole charge on the oxygen
atoms\cite{Zhang,Grant}. The $x^2 - y^2$ and $z^2$ orbitals are
degenerate in cubic symmetry (regular octahedron) but split, as shown
in Figure~2(a), when the octahedron is elongated along the $c$-axis
and contracted in the basal plane.  Such a tetragonal distortion,
Q$_{\theta}$, equally raises/lowers the energy of the $x^2 - y^2$ and
$z^2$ orbitals, respectively, by $VQ_{\theta}$, where $V$ is the
corresponding constant of electron-lattice coupling.  Due to the
different electron occupancies of these orbitals the decrease in
energy is $-VQ_{\theta}$.

The distortion accompanying the lifting of the orbital degeneracy is
known as the Jahn-Teller (JT) effect.\cite{Jahn-Teller,Bersuker} The
magnitude $Q_{\theta}$ of the distortion is determined by minimizing
the sum of the JT destabilizing energy, $-VQ_{\theta}$, and the
stabilizing energy of the resultant elastic strain,
$k{Q_{\theta}^2}/2$ ($k$ is the elastic force constant). At the
equilibrium distortion $Q_0$ the JT effect lowers the energy by
$-k{Q_0}^2/2 = -V^2/2k $.

\begin{figure}
 \psfig{file=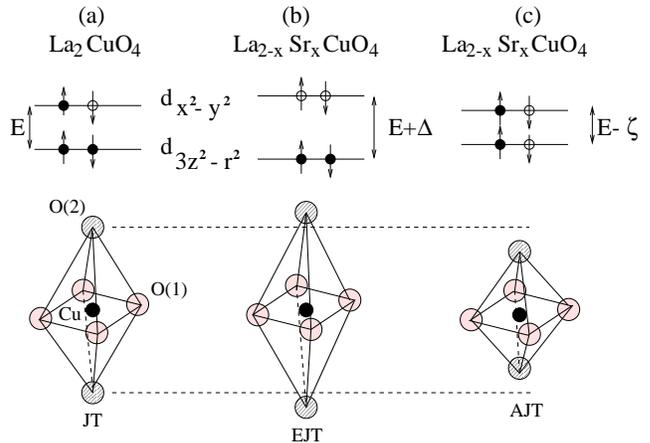,height=2.3in}
  \vspace{0.1truein} 
 \caption{Schematic electron energy levels and
corresponding distortions of CuO$_6$ octahedron due to
presence of intrinsic and extrinsic holes. (a) Tetragonal Jahn-Teller
distortion in pure La$_2$CuO$_4$ induced by intrinsic hole alone. (b)
Spin-singlet case, hole-induced enhancement of the Jahn-Teller effect,
and (c) Spin-triplet case, the anti-Jahn-Teller effect.}
\end{figure}

Figures~2(b) and 2(c) show the impurity energy levels for the
Sr-octahedrons with their two holes (intrinsic, extrinsic). The
situation of both holes in the $x^2 - y^2$ state with opposite spins
(Figure~2(b)), the singlet $^1A_{1g}$ state, is analogous to the
Zhang-Rice singlet\cite{Zhang} (two important differences are
discussed below). Figure~2(c) shows energy levels for holes in each of
$x^2 -y^2$ and $z^2$ orbitals. In this configuration, the triplet
$^3B_{1g}$ state, the holes have parallel spins and their exchange
(Hund's rule) interaction lowers the triplet state energy to near
that of the singlet state.
    
In the triplet state no JT force is present since equal populations of
$x^2 - y^2$ and $z^2$ orbitals result in no energy gain induced by a
tetragonal distortion.  This results in hole-induced contraction of
the Sr-octahedrons, called the {\it anti-JT}
effect\cite{Bednorz,Anisimov}. In the singlet state,
however, the JT energy gain is $-2VQ_{\theta}$, twice that of the
La-octahedrons resulting in an {\it enhanced JT} elongation of the
Sr-octahedrons (twice as big).

The {\it anti-JT} and {\it enhanced JT} distortions of Sr-octahedrons
in \lasco explain the double site observed by XAFS for the apical
oxygen that bridges the Cu and Sr atoms.  In the singlet state ({\it
enhanced JT} case) the apical oxygens are shifted away from Cu towards
the Sr impurity atom, producing the short Sr-apical oxygen distance.
Similarly, the triplet state ({\it anti-JT} case) shifts the apical
oxygen towards Cu away from the Sr atom, producing the long Sr-apical
oxygen distance. Thus, the two sites detected by XAFS give direct
evidence for two kinds of spin-differentiated JT distortions induced
by the donated holes.
    
The above theoretical arguments predict equal but opposite
displacements of the two sites from the La-apical distance of
2.35$\rm\, \AA$. The experiments find that the enhanced JT site is
displaced by 0.1$\rm\, \AA$ while the anti-JT site is displaced by
0.2$\rm\, \AA$ (in opposite directions). As is customary in discussing
JT distortions, these arguments assume a symmetric environment of the
Sr-octahedron and neglect its breathing mode contraction induced by
the extrinsic hole. Having a Sr dopant at one end and a La atom at the
other breaks the symmetry and, together with the breathing mode,
modify the distortions.

The change in energy due to the JT distortions can be estimated by
$E_{JT}\sim -kx_0^2$, where $x_0$ is the displacement of each of the
two apical oxygens from the configuration with no JT driving force.
Here the approximation is used that since the apical oxygens distort
much more than the planar oxygens, their contribution to the energy
change dominates. The anti-JT triplet has no driving force from the JT
effect and will be used as the zero of the displacement and JT energy
(its associated octahedron, however, is not regular due to the
tetragonal field of the average crystal structure). The XAFS
measurements determine $k=2.6$ eV/$\rm\, \AA^2$ from the temperature
dependence of the Cu-apical bond distance and $x_0$ is determined
directly from the two sites around the Sr. The values of $E_{\rm JT}$
and $x_0$ are listed in Table~\ref{tab:1} for the three states:
triplet, intrinsic or La-octahedron, and singlet. In the case of the
singlet state the {\it enhanced JT} distortion lowers the energy
relative to that of the intrinsic La-octahedron.

The ionization energy to remove the local extrinsic hole from the
singlet or triplet state into the extended state of the conduction
band includes the JT distortion energy and the local hole-impurity
interaction energy. The lack of temperature dependence in the relative
occupation of singlet and triplet states found by XAFS, up to RT,
indicates their binding energies are larger than 300 K$_B\approx 25$
meV. This is in agreement with LDA + U first principles
calculations\cite{Anisimov} that include the impurity potential at
$x=0.25$ but not allowing relaxation of the lattice. The calculation
finds that doped holes occupy impurity states with $x^2-y^2$ symmetry
located inside the Mott-Hubbard gap with an ionization energy of $\sim
0.1$ eV. Relaxing the lattice from the JT to the enhanced JT value
lowers the energy further by $\sim 0.03$ eV. The same paper
calculates, neglecting the impurity potential but including the
lattice relaxation\cite{Anisimov}, the triplet state to lie only
$\approx 54$ meV above the ground state. It is reasonable to expect
that including the Sr impurity potential will lower the triplet to
overlap the singlet band, resulting in both being populated by the
doped holes, as found in our measurements. Note from Table~\ref{tab:1}
that the enhanced JT energy (-0.23 eV) is an important contribution
for stabilizing the singlet impurity state.

The large binding energy $\gtrsim$ 0.1 eV of the impurity states pairs
is consistent with polarized x-ray absorption near edge structure
(XANES) measurements at RT\cite{Chen-Pellegrin} on the O $K$
edge. These show a large spectral weight transfer of the oxygen $2p$
hole states, with doping, from those in the upper Hubbard band
(intrinsic holes) to the impurity hole states introduced with
doping. This weight transfer, which occurs even at RT, shows the
impurity states are an intimate mixture of intrinsic and extrinsic
holes due to the strong interaction between them, consistent with
their strong pairing in our model.

The interaction between the Sr-impurity and its extrinsic hole
introduces binding which is essential for producing the impurity
states. The impurity states are formed because it is energetically
favorable to keep the extrinsic holes near the impurity than have them
distributed periodically along the Cu-O planes.  The singlet impurity
state is different from the Zhang-Rice singlet\cite{Zhang}, which is
periodically distributed over all of the CuO$_6$ octahedra. Another
important difference is that both the singlet and the triplet impurity
states are accompanied by a considerable distortion of the
Sr-octahedrons from the undoped configuration.
    
At a concentration $x_c \approx 0.07$ the impurity local energy levels
transform into impurity bands and the material becomes conductive. The
question arises which impurity state dominates the conductivity, the
singlet or the triplet. Just above $x_c$ the answer is given by that
state which has the greatest overlap between neighboring Sr sites as
the other will still be localized.  The singlet state, having a
greater extent in the plane, has a greater overlap integral by a
factor of three than the triplet\cite{Khomskii,Zaanen} at a given $x$,
if the planar amplitude of the states is the same. Our cluster
calculations indicate that this factor may be significantly higher
since the $z^2$ triplet state has less probability to reside on the
CuO$_2$ planes than the $x^2-y^2$ state. Thus, the singlet state
dominates the conductivity above $x_c$, and probably does so
throughout the range where high T$_c$ exists.

We argue that the singlet and triplet states are present
simultaneously on each Sr-octahedron and not separately on different
ones. The latter case implies that there are two types of
Sr-octahedrons, $S$ and $T$, containing only singlet or triplet pairs,
respectively. Then two separate impurity ``bands'' would be formed,
one a singlet from the $S$ sites and the second a triplet from the $T$
sites. However, neither could conduct. The $S$-''band'' has two holes
per orbital state and thus is completely full while in the
$T$-''band'' the two orbital states are half full and would
have their own Mott-Hubbard gap. Having overlapping singlet and trilpet
states at each impurity site is critical for allowing impurity band
conduction. The overlap allows hole transfer between bands so that
each is neither half nor completely full.

The tendency of extrinsic holes to occupy $z^2$ orbitals was predicted
by Khomskii and Neimark\cite{Khomskii}. Zaanen {\it
et~al.}\cite{Zaanen} treated the two types of carriers more
extensively, including band formation, and obtained a concentration
dependence for them similar to the one in Fig.~1. The predominant
occupation of triplet states by the doped holes is in apparent
contradiction with the interpretation of XANES measurements by Chen
{\it et~al.} \cite{Chen-Pellegrin} of a predominant in-plane character
for the doped holes. However, as pointed out by Chen {\it et~al.},
concluding on the orbital character of doped holes from XANES
measurements is difficult due to the large spectral weight transfer
with doping from upper Hubbard band states to the impurity states, as
discussed above. This large weight transfer is contrary to the rigid
band model and shows a strong modification of the intrinsic holes
states due to the addition of doped holes. We note that the
measurements by Chen {\it et~al.} do find a large increase in the oxygen
$p_z$ orbital character of holes with doping, consistent with the
significant increase of $z^2$ states found by XAFS.

Although there is a JT distortion accompanying the singlet pair, this
is not a polaron. A polaron carries its distortion with it as it moves
from site to site. In the case here the distortion is at each impurity
site only, and does not follow the pair as it translates from site to
site as then the distortion would have to appear on the intervening
La-octahedrons.  The JT distortion just changes the local environment
about each Sr- octahedron only, with its attendant lowering of the
energy of the singlet state, as occurs for impurity states embedded in
a host of La-octahedrons.  Though the impurity state JT pairs discussed
here are bosons, they are not the bosons composed of bipolarons that
have been suggested as a possible mechanism for high-Tc
superconductivity\cite{Alexandrov-Mott}.

The strong self-binding of the singlet JT pairs suggests the
speculation that they may move in the crystal as an entity and form a
Bose-Einstein condensate with superfluid characteristics at low
temperature assuming an appropriate dispersion curve. Therefore, high
T$_c$ of \lasco would be explained as superfluidity of charged
bosons\cite{Micnas} where disappearance of superconductivity above
T$_c$ is not related to a dissociation of pairs, but to a loss of
coherence between pairs. Some additional arguments supporting this
possibility are the larger single particle excitation gap than the
Andreev gap\cite{Deutscher} and the persistence of a pseudogap above
T$_c$\cite{Renner}.
    
The two inequivalent copper sites found in nuclear magnetic and
quadrupole resonance (NMR, NQR)
measurements\cite{Yoshimura-Kumagai,Raivo-Hammel}, one of them
increasing with $x$, can be explained in our model by the presence of
La-and Sr-octahedrons. Coupling of singlet and triplet states via
perturbations such as spin-orbit and electron-electron interactions
mixes these states at a rate much faster than the inverse linewidth of
the NMR/NQR lines ($\sim 1 \mu$sec), giving an average signal for the
Sr-octahedrons.

In summary and conclusion, two coexisting, spin differentiated, JT
distortions of Sr-octahedrons are responsible for the double site
distribution of apical oxygens around Sr impurities. Intrinsic and
doped holes are paired in impurity states bound on Sr-octahedrons by
these lattice distortions and the attraction of the doped
hole to the Sr impurities.  The triplet states with out of plane
$z^2$ character increase their fraction of doped holes with $x$; yet
planar $x^2-y^2$ singlet states dominate the conductivity due to their
greater overlap. The high binding energy of singlet states, $\gtrsim$
0.1 eV, suggests they retain their integrity as they conduct and can
Bose-condense into a superconducting ground state. Since pairing
persists up to at least RT, the properties of the carriers above T$_c$
cannot be treated by a rigid band model, which assumes that intrinsic
and extrinsic holes are uncoupled, as is done when extrapolating band
calculations of the undoped material to describe the normal state
carriers.

It should be emphasized that the model proposed here has experimental
support only for \lasco and further experiments need to be performed
to determine how general this model is for understanding high T$_c$ in
other materials.

We are pleased to thank I. Bersuker, Y. Yacoby, G. Bersuker and
N. N. Gorinchoi for helpful and stimulating discussions, as well as
      the support of DOE Grant No.~DE-FG03-98ER45681.

\begin{table}[tb]
\caption{Jahn-Teller energies, $E_{\rm JT}$. $x_0$ is the
displacement of each apical oxygen from the Cu induced by the JT
effect. The Anti-JT, JT, and Enhanced JT octahedrons are ones with
triplet paired holes, an intrinsic hole, and singlet paired holes,
respectively.}\label{tab:1}
 \begin{tabular}{c c c c} 
{\rm Octahedron type} & {\rm Anti-JT} & {\rm JT} & {\rm Enhanced JT}
\\ \hline 
 $x_0(\rm\, \AA)$ & 0.0 & 0.2 & 0.3  \\ \hline
 $E_{\rm JT}$(eV)  & 0.0 & -0.10 & -0.23  \\ 
\end{tabular}
\end{table}

\end{document}